%% file: main.tex
\documentclass[10pt,conference]{IEEEtran}

\usepackage{cite}
\usepackage{amsmath,amssymb,amsfonts}
\usepackage{algorithmic}
\usepackage{graphicx}
\usepackage{textcomp}
\usepackage{xcolor}
\usepackage{xspace}
\usepackage{fontawesome}
\usepackage{multirow}
\usepackage{rotating}
\usepackage{tikz}
\usepackage{soul}
\usepackage{makecell}
\usepackage{url}
\usepackage{enumitem}

\newcommand{\ie}{\emph{i.e.,}\xspace}
\newcommand{\eg}{\emph{e.g.,}\xspace}

\newcommand{\etal}{\emph{et~al.}\xspace}
\newcommand{\secref}[1]{Section~\ref{#1}\xspace}

\newcommand{\figref}[1]{Fig.~\ref{#1}\xspace}

\newcommand{\tabref}[1]{Table~\ref{#1}\xspace}

\newcommand*\circled[1]{\tikz[baseline=(char.base)]{
		\node[shape=circle,draw,inner sep=1pt] (char) {#1};}}

\newcommand{\approach}{{FeaRS}\xspace}
\newcommand{\totalApps}{{20,713}\xspace}
\newcommand{\totalCommits}{{2,721,800}\xspace}
\newcommand{\usefulCommits}{{841,995}\xspace}
\newcommand{\nodes}{{2,018,479}\xspace}

\usepackage{ifthen}
\newboolean{showcomments}
\setboolean{showcomments}{true}
\ifthenelse{\boolean{showcomments}}
{\newcommand{\nb}[2]{
		\fbox{\bfseries\sffamily\scriptsize#1}
		{\sf\small$\blacktriangleright$\textit{#2}$\blacktriangleleft$}
	}
}
{\newcommand{\nb}[2]{}
}

\AtBeginDocument{%
  \providecommand\BibTeX{{%
    \normalfont B\kern-0.5em{\scshape i\kern-0.25em b}\kern-0.8em\TeX}}}

\begin{document}

\title{Siri, Write the Next Method}

\author{
\IEEEauthorblockN{Fengcai Wen, Emad Aghajani, Csaba Nagy, Michele Lanza, Gabriele Bavota}
\IEEEauthorblockA{\textit{Software Institute -- USI Universit\`{a} della Svizzera italiana, Switzerland}}
}

\maketitle

\input{abstract}

\begin{IEEEkeywords}
Code Recommender, Empirical Software Engineering, Mining Software Repositories
\end{IEEEkeywords}



\input{introduction}
\input{approach}
\input{design}
\input{results}
\input{threats}
\input{related}
\input{conclusion}

\section*{Acknowledgment}
We gratefully acknowledge the financial support of the Swiss National Science Foundation for the projects PROBE (SNF Project No. 172799) and CCQR (SNF Project No. 175513).

\bibliographystyle{IEEEtran}
\bibliography{IEEEabrv,references} 

\end{document}

%% file: abstract.tex

\begin{abstract}

Code completion is one of the killer features of Integrated Development Environments (IDEs), and researchers have proposed different methods to improve its accuracy. While these techniques are valuable to speed up code writing, they are limited to recommendations related to the next few tokens a developer is likely to type given the current context. In the best case, they can recommend a few APIs that a developer is likely to use next. We present \approach, a novel retrieval-based approach that, given the current code a developer is writing in the IDE, can recommend the next complete method (\ie signature and method body) that the developer is likely to implement. To do this, \approach exploits ``implementation patterns'' (\ie groups of methods usually implemented within the same task) learned by mining thousands of open source projects. We instantiated our approach to the specific context of Android apps. A large-scale empirical evaluation we performed across more than 20k apps shows encouraging preliminary results, but also highlights  future challenges to overcome.

\end{abstract}

%% file: introduction.tex

\section{Introduction} \label{sec:intro}

Developing high-quality software while reducing time-to-market are two classical contrasting objectives in the software industry. This translates into the need for increasing the productivity of software developers, by lowering their learning curves when dealing with unfamiliar code, and by maximizing the quality of the code they write. In response to these needs, researchers have proposed recommender systems for software engineering, defined by Robillard \etal as ``applications that provide information items valuable for a software engineering task in a given context'' \cite{Robilliard:2014}.

Some recommender systems pursue a long-lasting dream of software engineering research: The (semi-)au\-to\-mat\-ic generation of source code. The goal of these tools is speeding up the implementation of new code. Code completion techniques are nowadays one of the killer features of IDEs \cite{MurphyKF06}. Researchers have proposed different methods to improve code completion accuracy and, more in general, its capabilities \cite{Bruch:fse2009,Hindle:icse2012,Nguyen:icse2012,Tu:fse2014,Robb2010a,Raychev:pldi14,Nguyen:msr2016}. While these approaches are certainly valuable to speed up code writing, they are limited to recommendations related to the next few tokens a developer is likely to type given the current context. In the best case, they can recommend a sequence of APIs that a developer is likely to use next \cite{Nguyen:icse2012,Raychev:pldi14}.

We aim at reaching the next level in supporting developers during the writing of new code. We present \approach, an approach and an IDE plugin which monitors the code written by Android developers in the IDE and is able to recommend the complete code of the next method (\ie signature and method body) they are likely to implement based on method(s) they already have implemented. 

\begin{figure}[!ht]
\centering
	\includegraphics[width=0.8\linewidth]{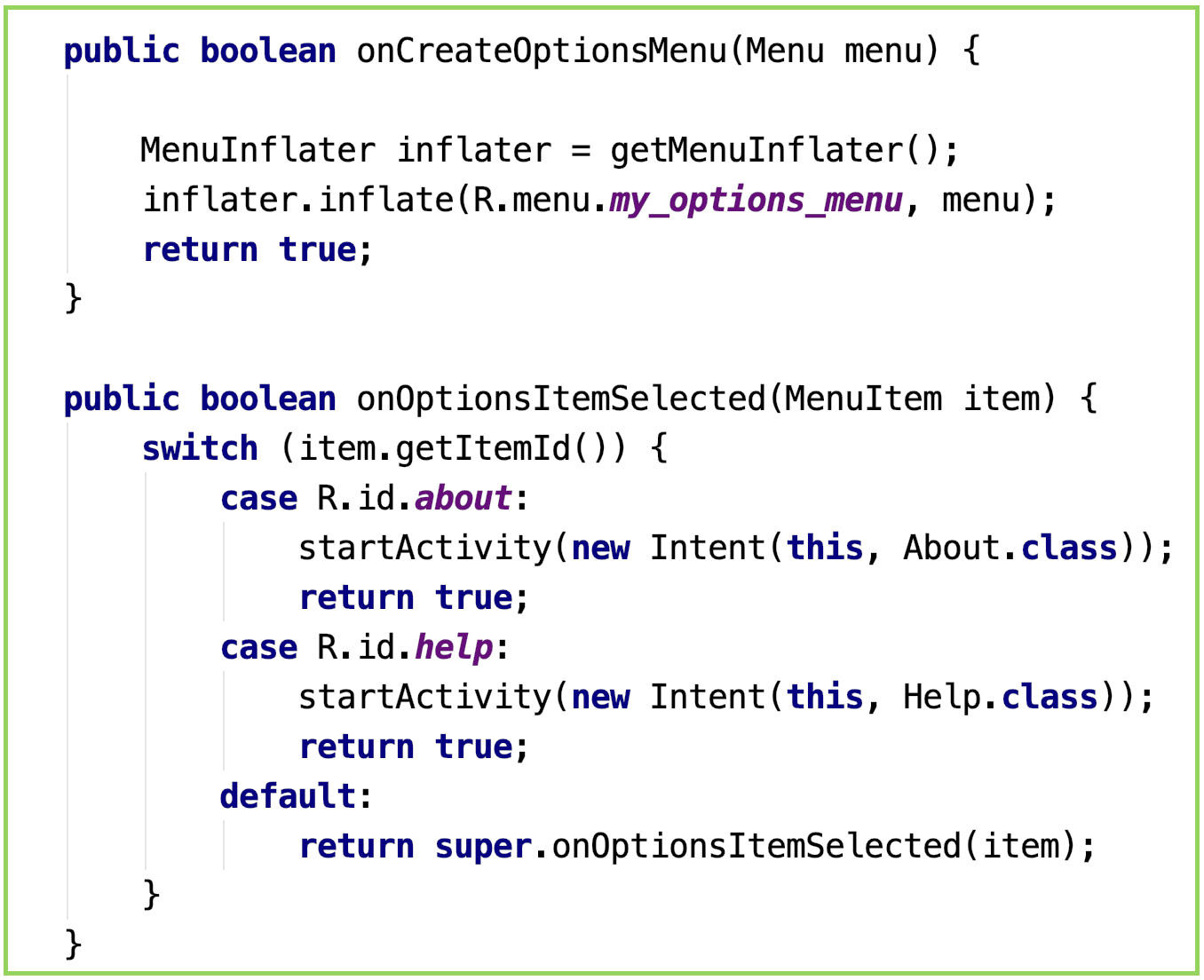}
	\caption{An implementation pattern in Android}
	\label{fig:motivation}
\end{figure}

\approach relies on a set of implementation patterns that we built by mining \totalApps open-source Android apps available on GitHub. To give a concrete example, the code snippet in \figref{fig:motivation} implements an options menu in an Android app. To perform such a task, tutorials recommend as first step to inflate the menu in the \texttt{on\-Create\-Options\-Menu(...)} method and, then, to handle the item selection in the \texttt{on\-Options\-Item\-Selected(...)} method. Assuming the existence of this implementation pattern in several apps, \approach can learn it and recommend the implementation of \texttt{on\-Options\-Item\-Selected(...)} once \texttt{on\-Create\-Options\-Menu(...)}  has been implemented by the developer.

We analyzed \totalCommits commits performed during the history of the subject apps to identify new methods that are implemented within the same commit. This results, for each analyzed commit $c_k$, in a set $M_k=\{m_1, m_2, \dots, m_n\}$ of $n$ new methods created in $c_k$. By extracting this information for thousands of commits, we can identify implementation patterns repeatedly followed by Android developers, \eg the implementation of $m_1$ could imply the implementation of $m_2, \dots, m_n$. We refer to $m_1$ as the Left-Hand Side (LHS) of the pattern and to $m_2, \dots m_n$ as the Right-Hand Side (RHS).

The identification of these implementation patterns is far from trivial. Indeed, two commits $c_k$ and $c_j$ performed in two different repositories may implement different sets of new methods (\eg $M_k=\{m_1, m_2\}$ and $M_j=\{m_3, m_4\}$) that, however, represent the same implementation pattern (\ie $m_1=m_3$ and $m_2=m_4$). Recognizing this situation is necessary to identify groups of methods that are repeatedly implemented together in different commits/apps, and not just by chance in a single/few commit(s).

\approach identifies clusters of methods likely to implement the same feature in the overall set of mined added methods. Going back to the previous example, this means that $m_1$ and $m_3$ are assigned to the same cluster $C_1$, and $m_2$ and $m_4$ to $C_2$. This results in the flattening of $c_k$ and $c_j$ to the same implementation pattern (\ie $M_k=M_j=\{C_1, C_2\}$). Once this processing is done for all mined commits, \approach applies association rule discovery \cite{agrawal1995mining} on all commits, thus creating the set of implementation patterns it relies on. 

When monitoring the code written by a developer in the IDE, \approach identifies newly written methods and assigns, if possible, each of them to one of the clusters created in the previous step. Then, it checks if an implementation pattern having one or more of the newly implemented methods as LHS is available and, in case a pattern is found, the corresponding RHS is triggered as a recommendation to the developer.

We evaluated \approach in a study in which we simulated its usage in the change history of the same \totalApps apps we used to extract the implementation patterns. We used the first 80\% of the apps' histories to extract the implementation patterns, the subsequent 10\% to tune the \approach's parameters, and the last 10\% to assess its performance (\ie test set). For each commit $c$ in the test set, we simulated the scenario in which a developer implemented a subset $S$ of the new methods added in $c$ and used \approach to generate recommendations using $S$ as LHS. Then, in case a recommendation is generated, we check if the RHS corresponds to one of the methods actually implemented in $c$ and not part of $S$. 

The achieved results show the feasibility of our approach, but also its strong limitations. Indeed, while \approach is able to generate meaningful recommendations for thousands of methods, several of them concern small methods that are not expected to substantially boost the developer's productivity.

%% file: approach.tex

\section{FeaRS} \label{sec:approach}

\begin{figure}[ht]
	\centering
	\includegraphics[width=\linewidth]{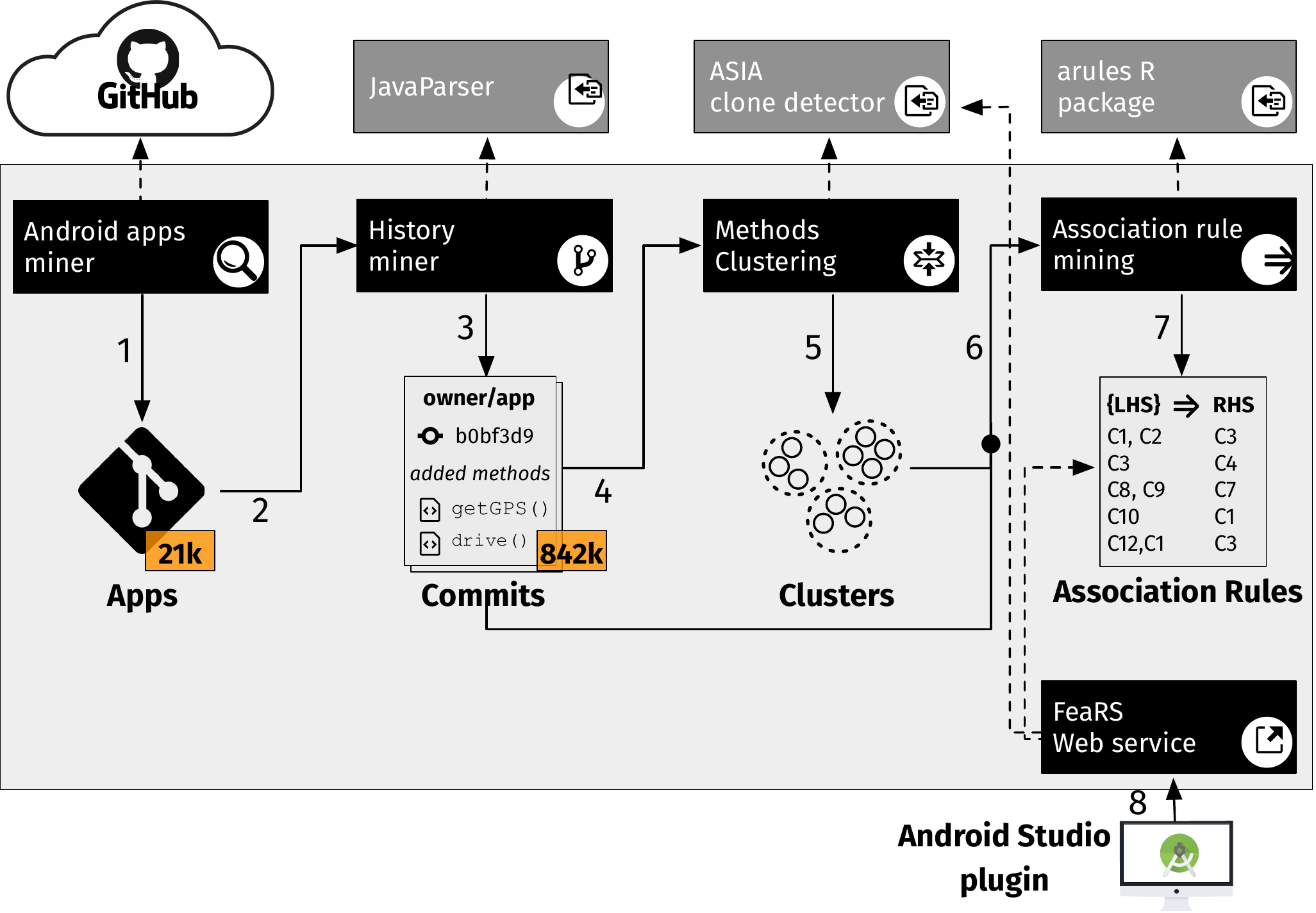}
	\caption{The \approach pipeline}
	\label{fig:approach}
\end{figure}

\figref{fig:approach} depicts the inner working of \approach. 

The black boxes represent components that we developed; the grey boxes depict external tools we reused and/or adapted. 

All components except the Android Studio IDE plugin reside on a central server providing an access point via the \emph{\approach Web service}. Steps 1-7 are executed offline and only once. Step 8 is executed every time the developer completes the implementation of a new method.

\subsection{Mining Android Apps}

The \emph{Android apps miner} identifies GitHub repositories related to Android apps. Their history is then analyzed to identify methods implemented within the same commit. We use the GitHub APIs to search for repositories satisfying the following criteria:

\emph{They are written in Java}. While Android is transitioning to Kotlin as the official language, the majority of Android apps is still written in Java \cite{coppola2019characterizing}. Note that while we instantiated \approach to the specific problem of recommending complete methods for Java Android apps, all the steps in \figref{fig:approach} can be customized to any programming language.

\emph{They are Android apps}. We ensure that the repository contains a \texttt{build.gradle} file with an explicit dependency towards the Android SDKs, indicating the usage of the Gradle build system, the default choice in Android Studio.

\emph{They have a limited, but non-trivial change history}. We excluded apps with {\em less than 100 commits} since we are interested in identifying the new methods added by developers within the same commit. Also, we excluded apps having {\em more than 1,000 commits}, since we do not want \approach to learn coding patterns peculiar only to a few apps.

The \emph{Android apps miner} identified and cloned \totalApps GitHub repositories, the set of apps that we use in this work, available in our replication package \cite{replication}. The set can be expanded by re-running the Android apps miner.

\subsection{Identifying Methods Added in Commits}

The set of cloned repositories is provided as input to the \emph{History miner} (step 2 in \figref{fig:approach}). This component extracts the list of commits performed in all branches of each repository by using the $\mathtt{git}$ $\mathtt{log}$ -{}-$\mathtt{topo}$-$\mathtt{order}$ command. This command allows analyzing all branches of a project without intermixing their history, avoiding unwanted effects of merge commits. 

\emph{History miner} uses JavaParser \cite{javaparser} to extract, from the Java files added or modified in each commit, the AST nodes which represent the callable declarations (\ie methods and constructors). In particular, we are interested in the callable declarations added in each commit. Commits not implementing at least two new methods and/or constructors are excluded at this stage, since we want \approach to learn implementation patterns in the form of $\{M\} \implies m_i$, where $M$ represents a set of one or more methods and $m_i$ a method that \approach can recommend based on the fact that the developer implemented $M$. Thus, assuming $M$ to be a singleton, at least two new methods must be implemented in a commit (\ie the one in $M$ and $m_i$) to make it useful for learning. We excluded commits adding more than 10 new methods (14\% of the total number of commits), since these are likely to be tangled commits not representative of any specific implementation pattern \cite{Herzig:msr2013}.

Overall, we processed \totalCommits commits, of which \usefulCommits were useful for building \approach (\ie those adding at least two new methods and no more than ten). These commits are provided as input to the module in charge of the methods clustering (step 4 in \figref{fig:approach}).

\subsection{Clustering Similar Methods}

To identify recurring implementation patterns in the considered commits, \approach applies clustering to group methods added in different commits, possibly from different systems, that implement equivalent or very similar functionalities. Two commits $c_k$ and $c_j$ performed in two different repositories may implement different sets of new methods (\eg $M_k=\{m_1, m_2\}$ and $M_j=\{m_3, m_4\}$) that represent the same implementation pattern (\ie $m_1=m_3$ and $m_2=m_4$). \approach can identify, through association rule discovery, that these sets of methods represent a repetitive implementation pattern. 

\approach builds a weighted undirected graph. Each method added in any of the commits is considered as a node. The weight on the edges connecting each pair of nodes represents the similarity between the two corresponding methods. To assess similarity we use the publicly available ASIA clone detector \cite{aghajani2019automated}, since it (i) is designed to capture the similarity between two Android methods; and (ii) returns as output an easily interpretable value from 0 (min similarity) to 1 (max). We customized the ASIA similarity algorithm in two ways.

First, in the original implementation all terms in the two methods to compare are lowercased before computing their textual similarity. This is suboptimal in \approach, since high precision in the identification of related methods is fundamental. 

Experiments revealed that the similarity of methods is artificially boosted by lowercase transformation: Given two methods $m_1$ and $m_2$, it happens that a term appearing in the name of $m_1$ (\eg~{\tt date}) is matched with the type of an object appearing in $m_2$ (\eg~{\tt Date}). By not transforming {\tt Date} to lowercase, the presence of these two terms does not influence positively the similarity between $m_1$ and $m_2$. 

Second, while ASIA uses tf-idf (term frequency-inverse document frequency) as a weighting schema for the terms during the textual similarity computation, we only employ term frequency, because we noticed that a single term appearing in both methods and having a very high idf (\ie being very rare in the corpus) can result in a high similarity between the two methods, even if they implement completely different features. This is especially true in small methods, due to the low number of terms present in them and the strong impact a single shared term can have on their similarity.

The graph we built contains \nodes nodes. We prune all edges with a weight below a threshold $\lambda$ ($\lambda$ will be tuned in our evaluation). This creates a set of disconnected subgraphs, each one representing a cluster of methods implementing strongly related functionalities. Within each subgraph (\ie cluster) we identify the cluster centroid: the method with the highest number of edges, which serves as representative for that cluster. The centroid is used later on by the \approach Web service when interacting with the IDE plugin.

\subsection{Association Rule Mining}

This module takes as input the list of commits generated by the \emph{History miner} and the clusters output of the previous step (step 6 in \figref{fig:approach}) and creates a text file reporting in each line a set of methods added in the same commit and in the same file, using the cluster they belong to. For example, assuming a commit adding three methods $m_1$, $m_2$, and $m_3$ to a file $F_i$, and those methods being assigned to clusters $C_{12}$, $C_{8}$, and $C_{71}$, respectively, a line $C_{12},C_{8},C_{71}$ will be added to the file. We decided to split methods added in the same commit but in different files to extract more ``cohesive'' association rules, and to avoid learning recommendations that span different files (\ie the developer is working on $F_i$ and \approach recommends a method to add in $F_j$). 

\approach analyzes the created file using Association Rule Mining \cite{Agrawal:1993} to identify implementation patterns, relying on  the $R$ \texttt{arules} package. We use the first 80\% of the apps' commits to extract the association rules, 10\% for tuning the parameters of \approach and 10\% to evaluate it. The output is a set of association rules in the form \{LHS\} $\implies$ RHS, where the LHS can be composed by one or more methods, while the RHS always has a single method. This means that \approach can only recommend the next method to implement given the one(s) already implemented by the developer.

There are three parameters that we tune in our evaluation: minimum {\em support} ($sup$), {\em confidence} for the mined rules ($con$), and maximum size of the LHS ($max_{LHS}$).

The support ($sup$) indicates how frequently a rule is observed in the dataset and, in our case, represents the percentage of analyzed commits that contains the specific rule.

The confidence ($con$) assesses how often a given rule is actually true in the dataset. Given a rule \{LHS\} $\implies$ RHS, it is computed as the number of commits implementing in the same file all methods in the LHS and RHS divided by the number of commits implementing the LHS in the same file (with or without the RHS). Finally, we also tune the maximum size of the LHS ($max_{LHS}$).

\subsection{The FeaRS Android Studio Plugin} \label{sub:plugin}

\begin{figure}[ht]
	\centering
	\includegraphics[width=6.41cm]{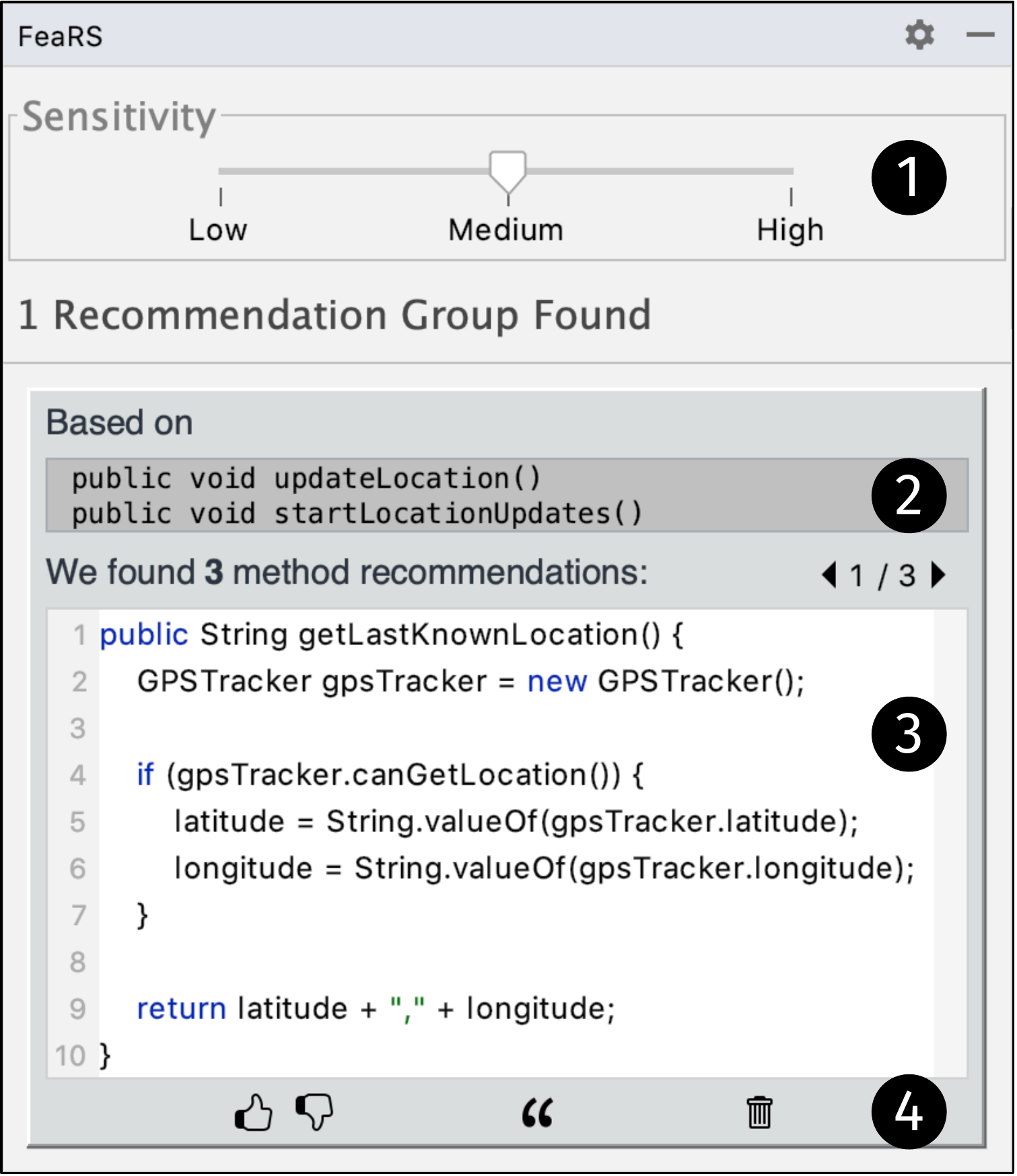}
	\caption{The FeaRS Android Studio plugin}
	\label{fig:plugin}
\end{figure}

\figref{fig:plugin} shows the \approach Android Studio IDE plugin.

The plugin interacts with the server through the Web service (step 8 in \figref{fig:approach}). The developer can start and stop \approach through simple \faPlayCircle{} and \faStop{} icons in the IDE toolbar. By clicking \faPlayCircle{}, \approach starts monitoring the code written by the developer and identifies when a new method is added. When this happens, the text of the new methods added by the developer since she pressed the start button is sent to the Web service. 

The Web service identifies, for each received method, the cluster it belongs to. Our customized version of the ASIA clone detector computes the similarity between each received method and each centroid representative of the computed clusters. The similarity $s$ for the most similar centroid is compared against a $\gamma$ threshold (the fifth and last \approach parameter to tune): If $s > \gamma$, the method is assigned to the cluster represented by the most similar centroid, otherwise no match is found and the method is discarded.

All combinations of received methods that are matched with a centroid are used to generate different LHSs. For example, if three methods added by the developer are matched to clusters $C_1$, $C_2$, and $C_3$, we generate 7 possible LHSs: \{$C_1$\}, \{$C_2$\}, \{$C_3$\}, \{$C_1,C_2$\}, \{$C_1,C_3$\}, \{$C_2,C_3$\}, and \{$C_1,C_2,C_3$\}. 

\approach checks if any of these LHSs is equal to the LHS of one of the association rules previously extracted. In case of a match, a recommendation is generated. In the reported example, if \{$C_1,C_2$\} is matched in a rule \{$C_1,C_2$\} $\implies$ $C_9$, then the centroid of cluster $C_9$ is returned by the Web service to the plugin as a recommendation. For the same LHS several different RHSs may be recommended. The matching of the LHS of two rules can lead to redundant recommendations. In the example, let us assume that two rules are matched, one with \{$C_1$\} and one with \{$C_1,C_2$\} as LHS, and that both of them have $C_9$ as RHS. In this case, the Web service returns the centroid of $C_9$ reporting that it is recommended based on the LHS belonging to the rule having the highest confidence.

The generated recommendations are shown in the IDE as depicted in the bottom part of \figref{fig:plugin}. \circled{2} shows the signatures of the methods implemented by the developer that are part of the LHS of the association rule used to recommend the method shown in \circled{3} (\ie RHS of the rule). In case several recommendations share the same LHS, the plugin displays them as one recommendation allowing developers to switch between different RHSs using the arrow buttons above \circled{3}. The buttons at the bottom of the code snippet \circled{4} allow to: (i) provide a feedback reporting if the recommendation was useful; (ii) copy the snippet; and (iii) delete the recommendation. The feedback, in our current implementation, is stored but not used. We plan to use it in future to adjust the confidence of the recommendations. If the developer decides to copy the snippet, a comment documenting the GitHub repository from when the snippet has been taken is added to the code, so that the developer can check its reusability from a legal perspective. 

The slider at the top of the plugin GUI \circled{1} allows the developer to customize the ``chattiness'' of the plugin on three different levels. \emph{Low}, \emph{Medium}, and \emph{High sensitivity} are three different \approach configurations that resulted from the calibration of its parameters presented in \secref{results_one}. By moving the slider towards \emph{Low}, \approach becomes more strict and generates fewer, but higher quality, recommendations, while the opposite holds for \emph{High}.

%% file: design.tex

\newcommand{\rqone}{What is the accuracy of \approach in recommending complete methods in the context of Android apps?}

\section{Study Design} \label{sec:design}

The goal of this study is to assess the performance of \approach when used to recommend the next method to implement given one or more (already implemented) methods as input. It thus addresses the following research question:

{\bf RQ$_{1}$:} {\em \rqone} 

\subsection{Context Selection and Data Collection} \label{sub:collection}

\begin{figure}[ht]
	\centering
	\includegraphics[width=0.7\columnwidth]{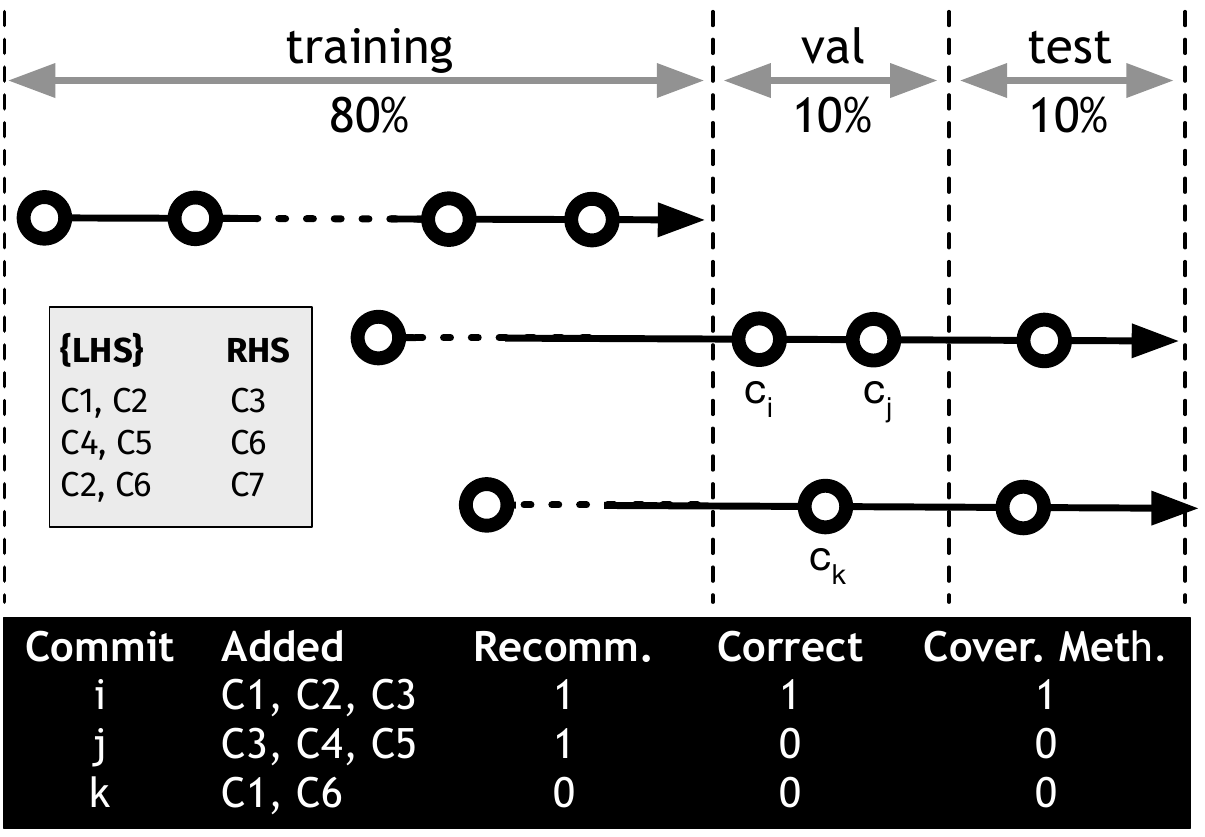}
	\caption{Data splitting and processing}
	\label{fig:evaluation}
\end{figure}

\begin{figure*}[ht]
	\centering
	\includegraphics[width=12.5cm]{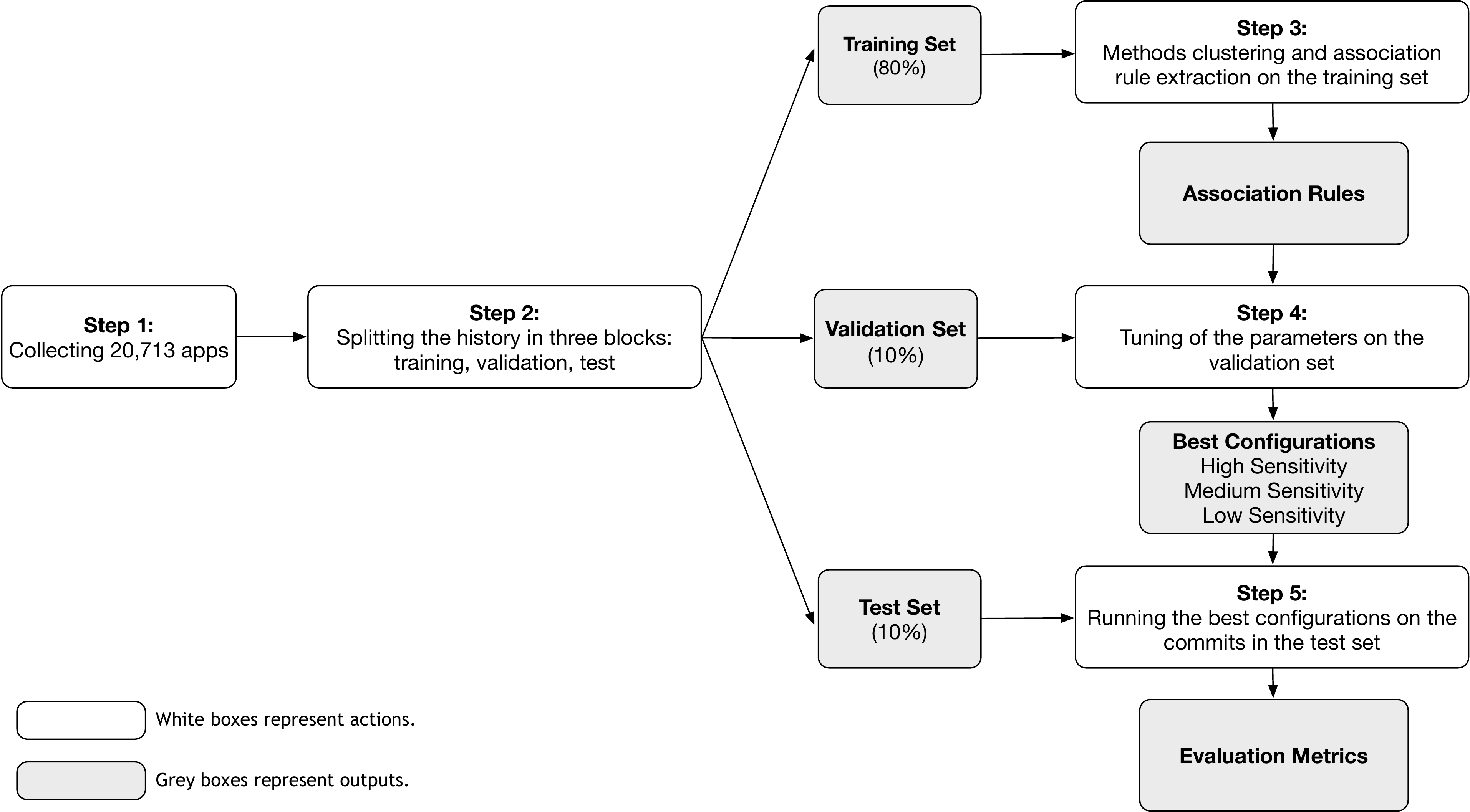}
	\caption{Study Design}
	\label{fig:design}
\end{figure*}

\figref{fig:design} overviews the steps in our experimental design. We exploit the dataset of \totalApps Android apps as the context of our study. Then, we split such a dataset into three blocks namely training, validation, and test. \figref{fig:evaluation} depicts how we create and use these three sets in our study.

The black arrows represent the change history of the apps considered in our study. Note that the history of the apps is not aligned, meaning that not all the apps exist in the same time period. The vertical dashed lines show how we divide the change history of the apps. 

We use the first 80\% to extract the association rules used by \approach to generate recommendations. We refer to this subset of the history as the ``training set.'' The subsequent 10\% is used to tune the parameters of \approach to identify the best configurations (\ie ``validation set''), which are used to generate recommendations on the ``test set'' (\ie the last 10\%), with the goal of assessing the performance of \approach.

One important clarification: We do not use the first 80\% of each repository as the training set, due to the misalignment of the mined change histories. Instead, given $d_s$ the date of the oldest commit present in all analyzed apps and $d_e$ the date of the most recent commit, we take the first 80\% of the time interval going from $d_s$ to $d_e$ as training set. As shown in \figref{fig:evaluation}, this may result in some apps exclusively contributing to the training set (or to the validation/test sets). 

However, such a design is needed to avoid using ``data from the future'' when generating recommendations for the validation and test set and, thus, to simulate a real usage scenario for \approach. Indeed, by selecting the first 80\% of the history of each app to learn the association rules, it could happen that a given $App_x$ has the last commit of training set made on date $d_x$, while for $App_y$ the latest commit of its entire history comes on date $d_y$, with $d_y < d_x$ (\ie $d_y$ is older than $d_x$). This would mean that association rules learned on $d_x$ will be applied to generate recommendations for commits performed on date $d_y$ (that will be part of the test set), thus using data from the future to learn how to trigger recommendations, something that cannot happen in a real usage scenario.

\begin{table}[ht]
	\centering
	\caption{\approach parameters tuning options}
	\label{tab:tuning}
	\begin{tabular}{ll} \hline
		\textbf{Parameter} &  \textbf{Experimented values}  \\ \hline
		$con$ & 0.05, 0.20, 0.35, 0.50, 0.65, 0.80\\
		$sup$ & 8.00E-06, 4.80E-05, 8.80E-05, 1.28E-04, 1.68E-04\\
		$\lambda$ & 0.80, 0.85, 0.90, 0.95\\
		$max_{LHS}$ & 1, 2, 3, 4, 5, 6, 7, 8, 9\\\hline
	\end{tabular}
\end{table}

Once the association rules are learned, we assess the performance of \approach on the validation set with different parameter configurations (\tabref{tab:tuning}), for a total of 1,080 configurations. Given the number of mined commits, the minimum value of $sup$ we experiment (\ie 8.00E-06) ensures that an association rule is learned from at least 5 commits to be considered valid. 

In all combinations of parameters, we used $\gamma = \lambda$, meaning that the minimum similarity needed to cluster two methods together (\ie $\lambda$) is also the minimum similarity used when generating recommendations to assign a newly implemented method $m$ to a cluster $C$ (\ie $\gamma$, see \secref{sub:plugin}).

As shown in \figref{fig:evaluation}, to identify the best configuration(s) we use 10\% of the apps change history (validation set). 

For each commit in the validation set ($c_i$, $c_j$, and $c_k$ in \figref{fig:evaluation}) we match all newly added methods to the clusters that have been defined during the association rules extraction from the training set (using the same similarity threshold as for the clusters definition). This means that we simulated the scenario in which each of the added methods is written by the developer in the IDE, and the \approach plugin checks if the added method can be matched with any of the existing clusters (\ie if its similarity with one of the centroids is higher than $\gamma$). If a method is not matched, no further action is taken, while all matched methods are assigned to the corresponding cluster. 

\figref{fig:evaluation} represents our running example, in which the grey box on the left shows the association rules learned on the training set, and the black box at the bottom shows how performance is computed for each commit in the evaluation set. In the case of commit $i$, three added methods have been matched to clusters $C_1$, $C_2$, and $C_3$. Then, we compute all possible combinations of the matched clusters involving all but one of them. In the case of commit $i$, this means all possible combinations having length lower than three: \{$C_1$\}, \{$C_2$\}, \{$C_3$\}, \{$C_1, C_2$\}, \{$C_1, C_3$\}, \{$C_2, C_3$\}. Then, we check if any of those combinations match the LHS of one of the rules learned from the training set. In \figref{fig:evaluation} the pair \{$C_1, C_2$\} matches the rule \{$C_1, C_2$\} $\implies$ $C_3$. This means that, assuming $C_1$ and $C_2$ to be written before $C_3$ (more discussion on this assumption in our threats to validity), \approach would be able in a real usage scenario to successfully recommend the next method to implement (\ie the $C_3$ centroid). Thus, in \figref{fig:evaluation}, we count the number  of recommendations generated by \approach (1), column ``Recomm.'', the number of correct recommendations (1), and the number of methods added in commit $i$ that \approach would have potentially been able to recommend (1 out of 3), column ``Cover. Meth.'' Concerning commit $j$, it would match the rule \{$C_4, C_5$\} $\implies$ $C_6$ generating one wrong recommendation (see \figref{fig:evaluation}). No recommendation would be triggered for commit $k$, since no matched rules are found.

There are two special cases that must be handled:

First, when multiple association rules have the same RHS (\eg assume \{$C_1$\} $\implies$ $C_3$ and \{$C_2$\} $\implies$ $C_3$ are both available in the set of learned association rules). In this case, both rules could be applied, for example, in the context of commit $i$ in \figref{fig:evaluation}. However, considering both rules as successful would inflate the performance of \approach since, in a real usage scenario, if \{$C_1$\} $\implies$ $C_3$ is applied, \{$C_2$\} $\implies$ $C_3$ cannot be applied, since $C_3$ already exists. 

Second, in case of a ``circular dependency'' between the LHS and the RHS of two rules, \eg $r_1$ = \{$C_1$\} $\implies$ $C_3$ and $r_2$ = \{$C_2, C_3$\} $\implies$ $C_1$. The LHS of $r_1$ matches the RHS of $r_2$, and the RHS of $r_1$ is contained in the LHS of $r_2$. 

In theory both rules could be applied to commit $i$ in \figref{fig:evaluation}, but the application of one rule would exclude the other in a real usage scenario. 
If we apply $r_1$, it means that $C_1$ has been implemented by the developer and it does not make sense to recommend it with $r_2$. Similarly, if $r_2$ is applied, this means that $C_3$ already exists, making $r_1$ useless.

In both cases we select the rule with the highest confidence.

\subsection{Data Analysis}

\begin{figure}[ht]
	\centering
	\includegraphics[width=0.9\columnwidth]{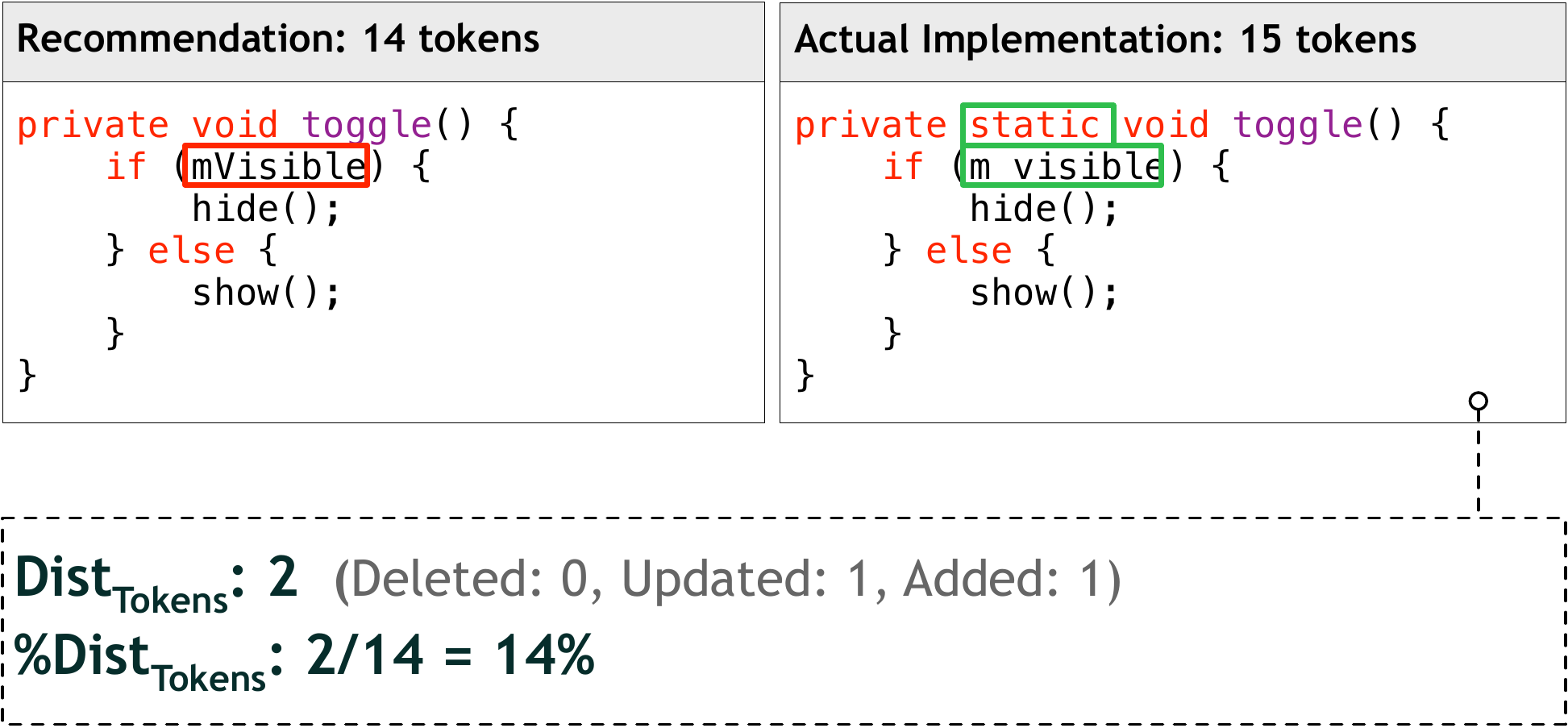}
	\caption{An example of $dist_{tokens}$ calculation}
	\label{fig:disttokens}
\end{figure}

We assess the performance of each experimented configuration by computing the following metrics:

\begin{description}[leftmargin=0pt,labelindent=\parindent] 

\item[Recall:] 
$recall=\frac{Comm_{cor}}{Comm_v}$, where $Comm_{cor}$ is the number of commits for which \approach generated at least one \emph{correct} recommendation and $Comm_v$ is the set of commits mined in the validation set. A correct recommendation is not necessarily an exact match to the actual implemented code, but the similarity has to be above a certain threshold which is consistent with the predefined clusters. Recall indicates in how many commits \approach could be potentially useful for developers.

\item[Precision:] 
$precision=\frac{Comm_{cor}}{Comm_{rec}}$, where $Comm_{rec}$ is the number of commits for which \approach generated at least one recommendation (correct or wrong).


\item[Cov\textsubscript{commits}:] 
$cov_{commits}=\frac{Comm_{rec}}{Comm_v}$. This metric indicates the percentage of commits from the validation set that could have triggered \approach to generate at least one recommendation (correct or wrong) for developers.

\item[Cov\textsubscript{meth}:] 
$cov_{meth}=\frac{Meth_{cor}}{Meth_{Comm_v}}$, where $Meth_{cor}$ is the number of methods successfully recommended by \approach and $Meth_{Comm_v}$ is the total number of methods added in $Comm_v$. This coverage metric indicates the percentage of methods added in all commits from the validation set that could have been automatically generated by \approach.

\item[\#Recom:]
 \#$recom$ is the number of recommendations generated by \approach in a commit for which it was triggered. We report both the mean and the median values. 

\item[Dist\textsubscript{tokens}:] 
$dist_{tokens}$ is the distance in number of tokens that must be modified, added or deleted by a developer when they receive a correct recommendation from \approach, which does not imply an exact match with the code actually implemented by the developer. Thus, we assess the effort needed by developers to adapt the received recommendation to their codebase (an example computation of such a metric is shown in \figref{fig:disttokens}).

\end{description}

\begin{figure*}[ht]
	\centering
	\includegraphics[width=12cm]{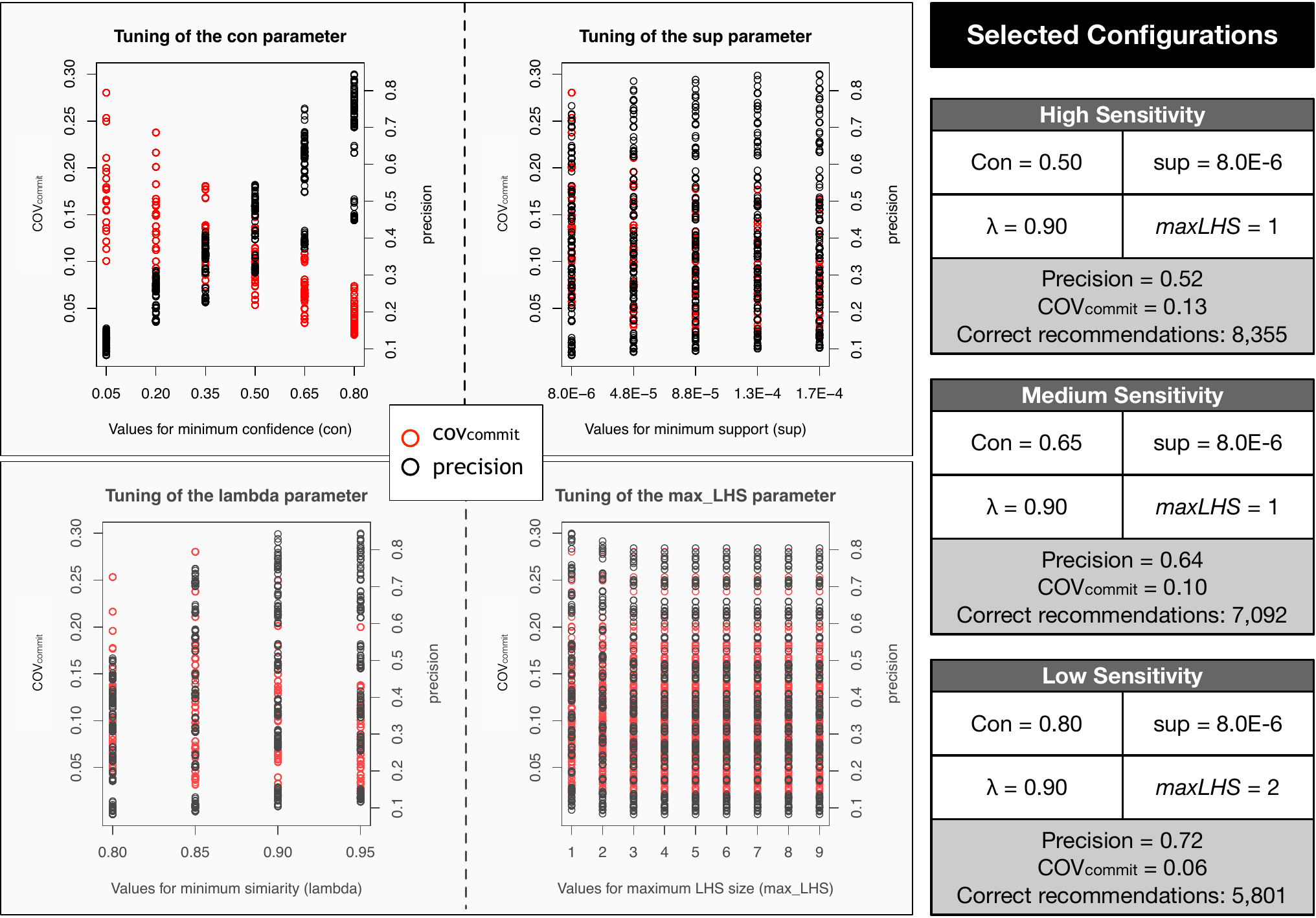}
	\caption{Tuning of \approach's parameters}
	\label{fig:tuning}
\end{figure*}

%% file: results.tex

\section{Results Discussion} \label{sec:results}


\input{results_one.tex}
\input{results_two.tex}
\input{results_three.tex}

%% file: results_one.tex

\subsection{\approach Parameters Tuning} \label{results_one}

\figref{fig:tuning} shows the results of the parameters tuning performed on the validation set. Each of the four graphs reports on the x-axis the values experimented for a specific parameter; from left to right: minimum confidence ($con$), minimum support ($sup$), minimum similarity to cluster two methods ($\lambda$), and maximum size of the LHS ($max_{LHS}$). The y-axis reports the $cov_{commits}$ (left) and the precision (right) achieved, with red dots indicating $cov_{commits}$ values, and black dots precision values. We decided to use these two metrics, over the others, for the parameters tuning since we wanted to contrast the talkativeness of our tool (\ie in how many commits it generates a recommendation) against the precision of the generated recommendations. To better understand what the black and red dots represent, consider the $con$ graph when its value is set to 0.05. The dots plotted in correspondence of this value represent the performance achieved when fixing $con=0.05$ and varying all other parameters.

One first observation is related to the range of performance achieved by different configurations: The $cov_{commits}$ varies from 0.02 to 0.28, while the precision from 0.08 to 0.84. While the values of $cov_{commits}$ may look low, it is important to note that the validation set includes 70,562 commits. 

The trends observed for the four parameters indicate that $con$ has the strongest influence on performance. When the minimum confidence needed to trigger a recommendation grows, as expected the precision linearly increases with a corresponding linear decrease of recall (left part of \figref{fig:tuning}). Setting $con$ lower than 0.50 does not ensure acceptable precision.

Concerning $sup$, increasing its minimum value does not substantially increase precision while having a strong negative effect on $cov_{commits}$. Low values of this parameter are preferable. Instead, increasing the $\lambda$ parameter results in a notable increase in precision, especially when moving from 0.80 to 0.90/0.95. In this case, 0.90 seems to be a good compromise, also considering the minor loss of $cov_{commits}$ as compared to lower values. Finally, the $max_{LHS}$ does not play a big role in the performance of \approach. As the output of this tuning process, we identified three configurations that we linked to the sensitivity bar in our IDE plugin and that are shown in the gray boxes at the right of \figref{fig:tuning}. 

These configurations have been picked using the following process. We started from the assumption that a precision level below 0.50 (\ie one out of two generated recommendations is correct) is not acceptable. Then, we picked as a \emph{high sensitivity} configuration the one ensuring a precision of at least 0.50 and having the highest $cov_{commits}$. This configuration is able to generate 8,355 correct recommendations in the validation set, with a precision of 52\%. Then, we increase the minimum acceptable precision by 10\%, identifying the configuration ensuring at least a 60\% precision with the maximum $cov_{commits}$. This resulted in the \emph{medium sensitivity} configuration, that can successfully recommend useful methods in 7,092 cases, with a precision of 64\%. Finally, a further increase of the precision level to at least 70\%, led to the identification of the \emph{low sensitivity} configuration, that can recommend 5,801 correct methods, with a precision of 72\%. These three configurations are the ones we experiment with.

%% file: results_two.tex

\subsection{Quantitative Results} \label{results_two}

\tabref{tab:resultAll} reports the results achieved by the three \approach's configurations on the test set. The top part of the table reports the raw data used to compute the performance metrics in the bottom part of the table. In the top part, while ``\#commits w. corr. recomm.'' indicates the number of commits with at least one correct recommendation, ``\#corr. recomm.'' represents the number of correctly recommended methods, possibly more than one per commit.

\begin{table}[ht]
	\centering
	\caption{Performance when considering all methods}
	\label{tab:resultAll}
	\begin{tabular}{lrrr} \hline
		  & \textbf{high} & \textbf{medium} & \textbf{low}\\
		& \textbf{sensit.} & \textbf{sensit.} & \textbf{sensit.}\\\hline
		\#commits                                         & 69,480              & 69,480                & 69,480             \\ 
		\#added methods                                   & 219,331             & 219,331               & 219,331            \\ 
		\#commits w. recomm.                  & 8,757               & 6,447                 & 4,116              \\ 
		\#commits w. corr. recomm.            & 4,878               & 4,167                 & 3,110              \\ 
		\#recommendations                                 & 14,642              & 9,996                 & 7,170              \\ 
		\#corr. recomm.                         & 7,383               & 6,183                 & 5,149              \\ \hline
		recall                                            & 0.07              & 0.05                 & 0.04              \\ 
		precision                                         & 0.50               & 0.62                 & 0.72              \\ 
		coverage$_{commits}$                       & 0.13               & 0.09                 & 0.06              \\ 
		coverage$_{meth}$                      & 0.03               & 0.03                 & 0.02              \\ 
		\#recom(median)                      & 1                    & 1                      & 1                   \\
		\#recom(mean)                         & 1.67              & 1.55                  & 1.74              \\ \hline
		distance$_{tokens}$(Q1,Q2,Q3)      & 0,1,2           & 0,1,2               & 0,1,2          \\
		distance$_{tokens}$(mean)        & 1.94               & 2.03                & 1.81           \\
		 \%distance$_{tokens}$(Q1,Q2,Q3)    &0,13,22            & 0,13,22                & 0,13,20         \\
		\%distance$_{tokens}$(mean)        & 14\%               & 14\%               & 13\%           \\ \hline
	\end{tabular}
\end{table}

The results achieved by the three configurations are in line with what we observed on the validation set: precision goes from 0.50 (\emph{high sensitivity}) to 0.72 (\emph{low sensitivity}), with recall moves in an inverse direction, decreasing from 0.07 (\emph{high sensitivity}) to 0.04 (\emph{low sensitivity}). 

The recall values, while low, still correspond to thousands of methods correctly recommended. As we learned while performing the qualitative analysis in \secref{results_three}, a correct recommendation does not imply a ``useful'' recommendation. We noticed that many of the correct recommendations are due to small methods (\eg a getter method triggers the implementation of the corresponding setter), and decided to re-compute the performance of \approach only considering recommended methods with at least four lines of code (including signature but excluding annotations and the closing brace).  To correctly compute recall, this also required us to exclude from our analysis the commits in which a successful recommendation would not be possible at all, due to the absence of newly implemented methods having at least four lines.

\begin{table}[ht]
	\centering
	\caption{Performance when excluding short methods}
	\label{tab:resultLong}
	\begin{tabular}{lrrr} \hline
		  & \textbf{high} & \textbf{medium} & \textbf{low}\\
		& \textbf{sensit.} & \textbf{sensit.} & \textbf{sensit.}\\\hline
		\#commits                                         & 31,088              & 31,088                 & 31,088              \\ 
		\#added methods                                   & 83,562             & 83,562               & 83,562            \\ 
		\#commits w. recomm.                    & 900               & 763                 & 564              \\ 
		\#commits w. corr. recomm.            & 568               & 536                 & 413              \\ 
		\#recommendations                                 & 1,329              & 1,099                 & 738              \\ 
		\#corr. recomm.                        & 778               & 742                 & 522              \\ \hline
		recall                                            & 0.02               & 0.02                 & 0.01              \\ 
		precision                                         & 0.59               & 0.68                 & 0.71              \\ 
		coverage$_{commits}$                      & 0.03               & 0.03                & 0.02            \\ 
		coverage$_{meth}$                      & 0.01               & 0.01                 & 0.01              \\ 
		\#recom(median)                      & 1                    & 1                      & 1                   \\
		\#recom(mean)                         & 1.48              & 1.44                  & 1.30              \\ \hline
		distance$_{tokens}$(Q1,Q2,Q3)      & 0,3,10           & 0,3,10               & 0,3,4         \\
		distance$_{tokens}$(mean)        & 5.08               & 5.07                & 3.98           \\
		 \%distance$_{tokens}$(Q1,Q2,Q3)    &0,14,28            & 0,14,28                & 0,10,18        \\
		\%distance$_{tokens}$(mean)        & 17\%               & 16\%               & 13\%           \\ \hline
	\end{tabular}
\end{table}

\tabref{tab:resultLong} reports the results achieved in this scenario. The precision values are in line with before (min: 0.59, max: 0.71), showing that the ``quality'' of the recommendations is not  influenced by the length of the recommended methods. Instead, we observed a drop of recall, that does not go over 2\%, with a number of correct recommendations ranging between 522 (low sensitivity) and 778 (high sensitivity). 

The number of recommendations generated by \approach (\#recom) is usually very low (median=1 and mean$<$2 in both scenarios). This shows that \approach does not generate many cases to inspect when triggered. Also, the results of distance\textsubscript{tokens} indicate that developers need to modify only a few tokens to adapt the received recommendations to their code. 

While these results show the potential of \approach, they highlight (as in cases discussed for \tabref{tab:resultAll}), that the recommended methods are short, with a potential small benefit for developers. Our qualitative analysis will help in better assessing the value of these recommendations. 

%% file: results_three.tex

\subsection{Qualitative Examples} \label{results_three}

\subsubsection{Correct Recommendations} 

\begin{figure}[ht]
	\centering
	\includegraphics[width=\linewidth]{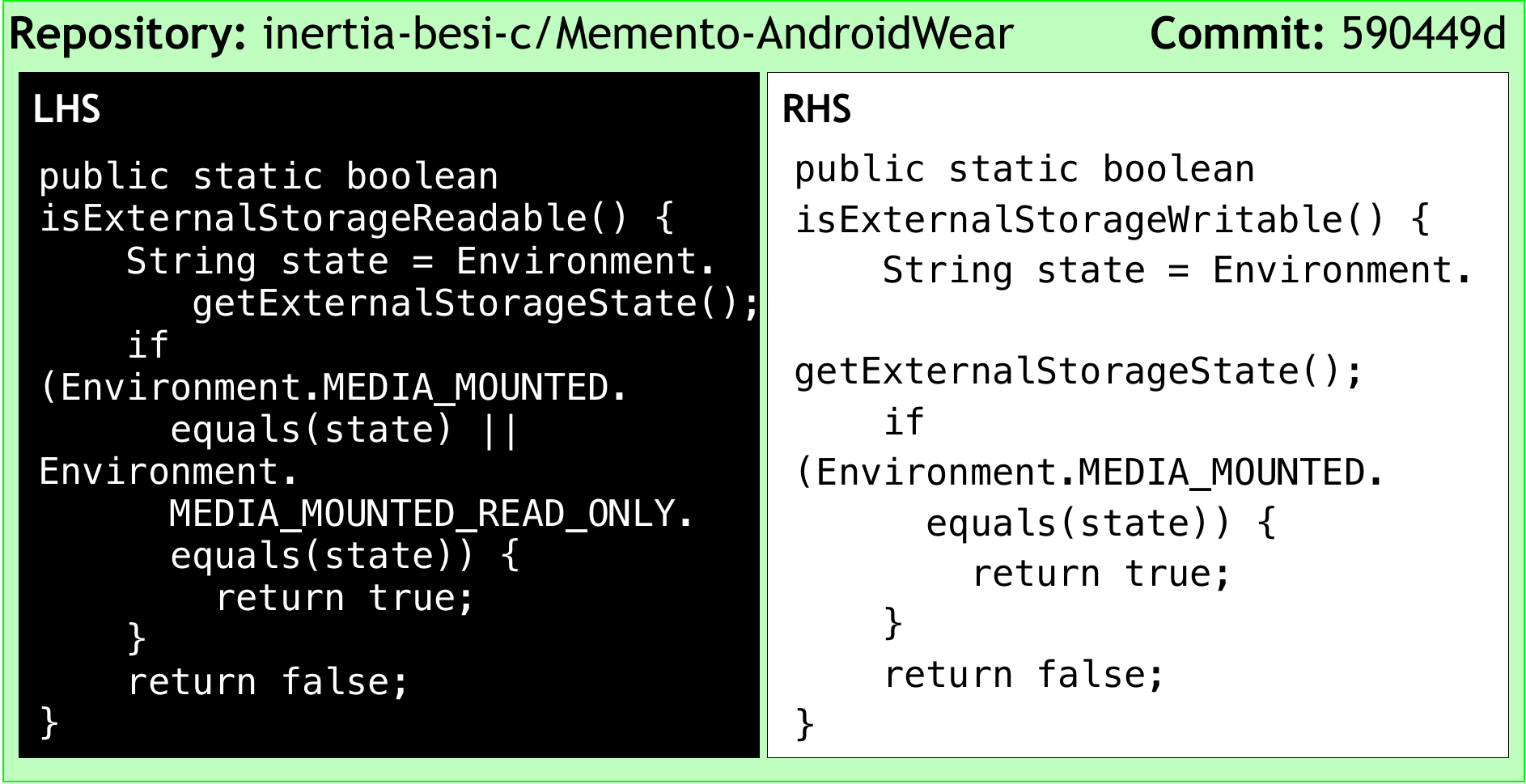}
	\caption{Correct recommendation to the usage of external storage in Android.}
	\label{fig:matched1}
\end{figure}

\figref{fig:matched1} shows an example of a recommendation generated for the Memento app for Android Wear\cite{memento}.

Suppose that the developer implements the \texttt{is\-External\-Storage\-Readable()} method to check whether the external storage of the device is mounted in read-only mode. \approach can pop up and recommend the \texttt{is\-External\-Storage\-Writable()} method to check also if it is writable or not. This rule had four matching instances in our test set from four different repositories.

\begin{figure}[ht]
	\centering
	\includegraphics[width=\linewidth]{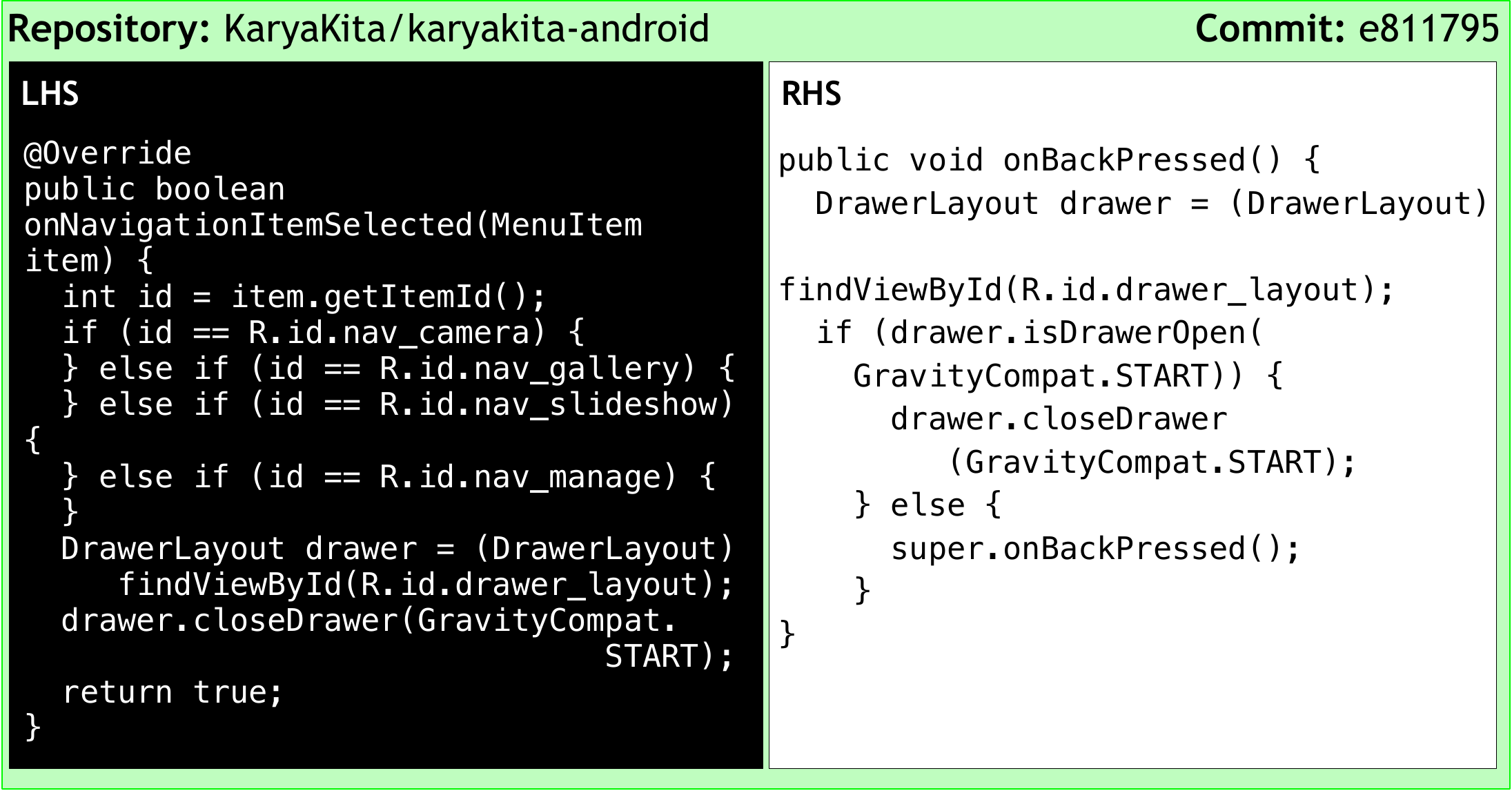}
	\caption{Correct recommendation to provide a custom back navigation for an Android DrawerLayout.}
	\label{fig:matched2}
\end{figure}

\figref{fig:matched2} shows an example of providing a custom back navigation for an Android {\tt DrawerLayout}.

Following the implementation of an \texttt{on\-Navigation\-Item\-Selected(...)} method that uses a \texttt{Drawer\-Layout}, \approach recommends a proper implementation for the \texttt{on\-Back\-Pressed()} method. Interestingly, in case of a missing implementation, the \texttt{Drawer\-Layout} might not close properly, as it is discussed in a Stack Overflow question\cite{stackoverflow26216088}. We found 19 matching instances for this rule in 17 different repositories.

\begin{figure}[ht]
	\centering
	\includegraphics[width=\linewidth]{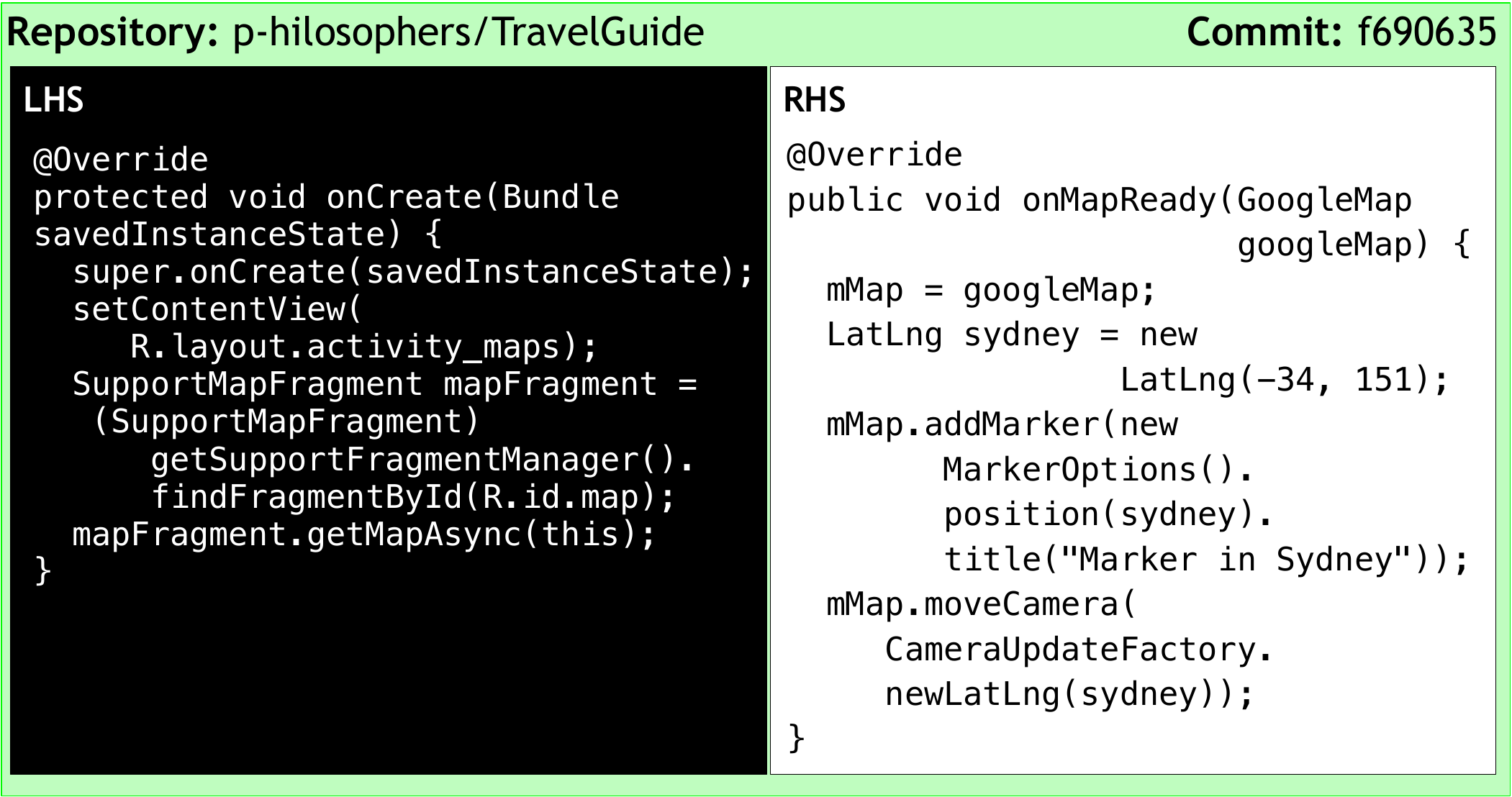}
	\caption{Correct recommendation for the creation of a GoogleMap instance from the Google Maps SDK for Android.}
	\label{fig:matched4}
\end{figure}

\figref{fig:matched4} shows an example recommendation for the creation of a Google Map object from the Google Maps SDK. 

We found 68 matches for this rule in 62 repositories. \approach matches an \texttt{on\-Create(...)} method in which an \texttt{Activity} creates a \texttt{Support\-Map\-Fragment} from the SDK. Next, it recommends an initial implementation for the \texttt{on\-Map\-Ready(...)} method, that shows how to add a marker to the map. We found various implementations having a different initial marker position (\eg London, Sydney).

\input{result.3.2.tex}

%% file: result.3.2.tex

\subsubsection{Unmatched Implementation Patterns}

We present \approach's recommendations that have been triggered during the evaluation process (\ie their LHS has been matched in the test commits) but that have never been successful (\ie the RHS has not been matched).

\begin{figure}[ht]
	\centering
	\includegraphics[width=\linewidth]{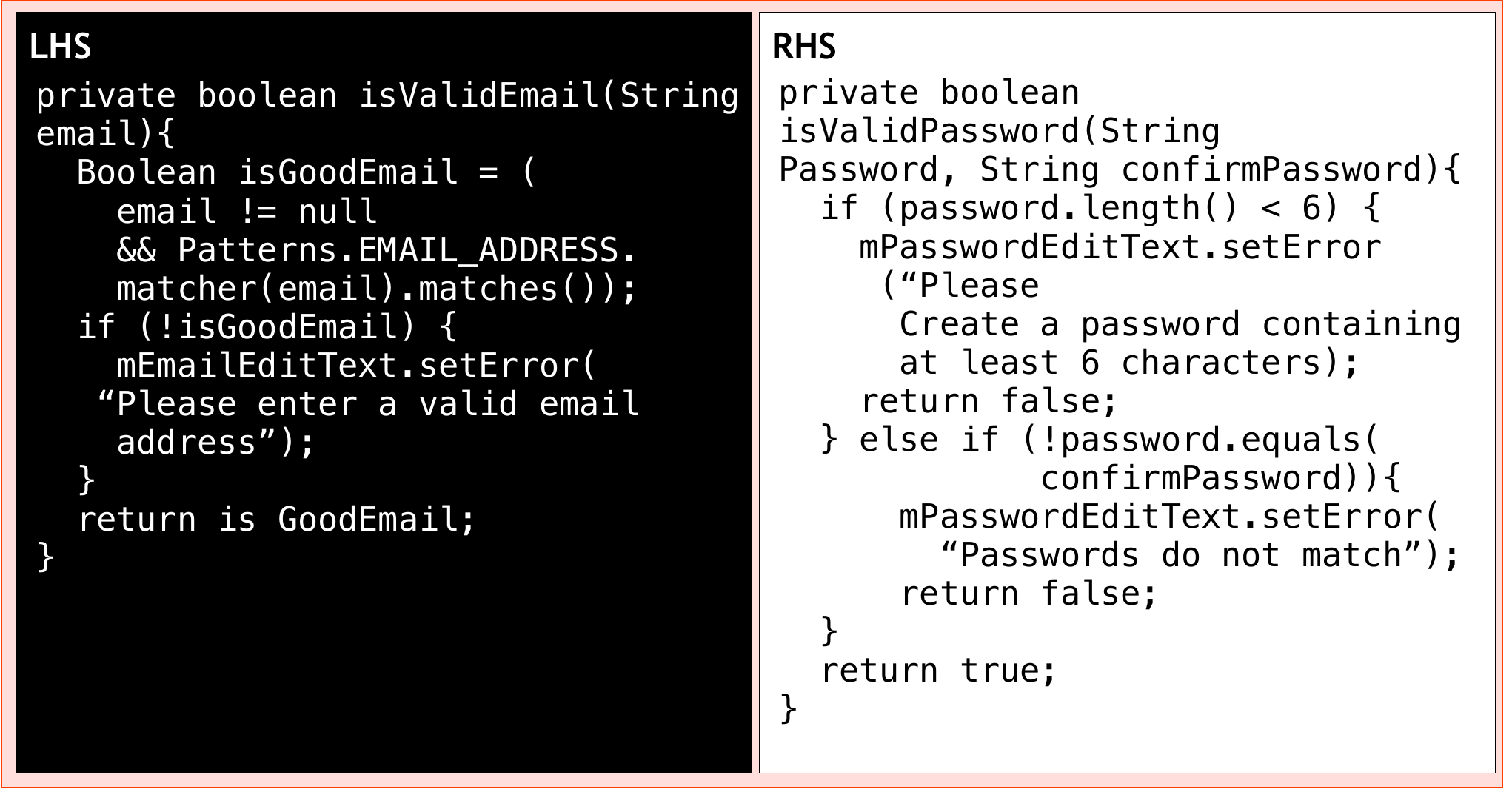}
	\caption{Unmatched recommendation for user credential validation in sign-up activity.}
	\label{fig:unmatched1}
\end{figure}

\figref{fig:unmatched1} shows an example of recommendation generated for the Artissans Android app\cite{artissans}. 

Suppose that the developer implements the \texttt{is\-Valid\-Email()} method to check whether the email address provided when creating an new account is valid. \approach recommends the \texttt{is\-Valid\-Password()} method to check, in the same scenario, if the provided password/confirm password fields are valid (\ie they are composed by at least six characters, and they match each other). This rule had been triggered twice without finding a match for the RHS, thus being classified as an incorrect recommendation. However, when we looked into the two commits in which this recommendation was triggered, we found that both of them actually implemented an \texttt{is\-Valid\-Password()} method that, however, only validated the password based on its length, do not making the recommended method and the implemented one similar enough to be counted as a correct recommendation. This example is representative of others we found.

\begin{figure}[ht]
	\centering
	\includegraphics[width=\linewidth]{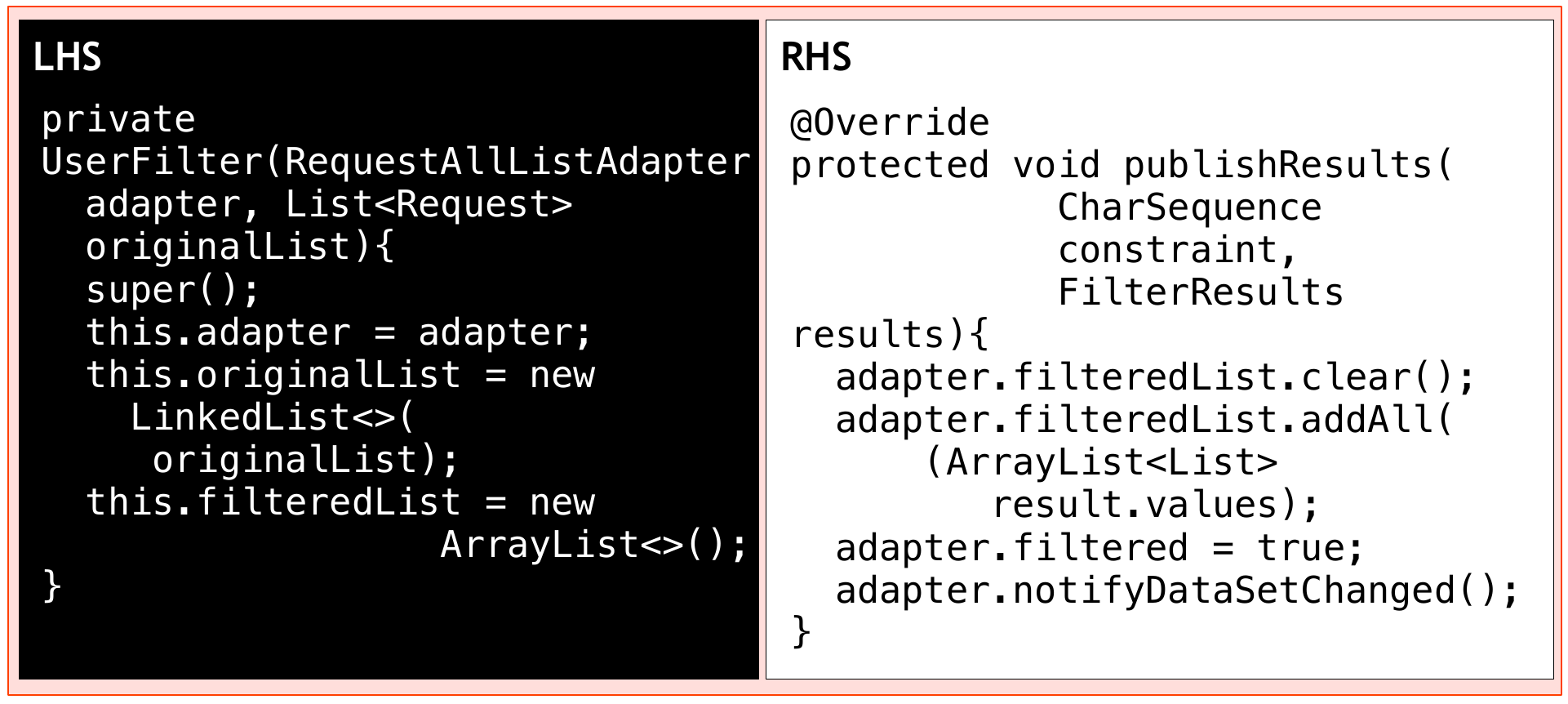}
	\caption{Unmatched recommendation for creating custom filter for filterable adapter in Android.}
	\label{fig:unmatched2}
\end{figure}

For example, \figref{fig:unmatched2} relates to the creation of a custom filter applied to a \texttt{Recycler\-View.\-Adapter} in Android. The class {\tt Filter} is used in Android to constrain data according to a specified pattern. 

Following the implementation of a \texttt{User\-Filter} constructor, \approach recommends a proper implementation of the overridden \texttt{publish\-Results} method from the \texttt{Filter} class that, as explained in the Android documentation, is \emph{invoked in the UI thread to publish the filtering results in the user interface}. Again, this recommendation was not matched (and considered wrong) during our study, but also in this case looking into the test commit\cite{artie} subject of the recommendation, we found that a similar overridden \texttt{publish\-Results} method was implemented as well following a custom filter constructor. Unfortunately, also in this case the similarity between the RHS of the rule and the implemented \texttt{publish\-Results} was not high enough to identify the recommendation as useful. 

These cases show that our experimental design, while useful to provide a first indication about the quality of the recommendations triggered by \approach, has imprecisions in assessing \approach's performance. As previously said, only complementing this mining-based study with experiments with developers can help in better assessing \approach's usefulness.

%

%% file: threats.tex

\section{Threats to Validity} \label{sec:threats}

\emph{Construct validity.} In our experimental design we assumed that if a commit added three methods belonging to clusters $C_1$, $C_2$, and $C_3$ and \approach has an association rule \{$C_1$\} $\implies$ $C_3$, \approach would have been useful in that commit to recommend $C_3$ to the developer. However, we cannot know whether $C_3$ was written before $C_1$, thus making \approach's recommendation useless in practice. Such a threat can only be addressed by (i) performing a user study in which developers code live using \approach, or (ii) recording IDE interaction data of programming sessions. While this is part of our future work, we preferred as first evaluation for \approach something that can be large-scale and fully automated, before moving to more costly studies requiring human involvement. In the design of our study, we only consider coding activities from one single commit might perform an implementation task, while ignoring those cases in which a given task can be separated into several commits. Actually we considered the idea of using close commits as a single data point, but we found out that it is hard to define a proper criterion for the selection of multiple commits and it might be risky for the cohesiveness of the task.

Another threat is related to the criterion we used to identify a generated recommendation as ``correct.'' Given a commit $c$ in which $m_i$ and $m_j$ are added, we assume that a recommendation $C_k$ $\implies$ $C_s$ is correct if $m_i$ is matched to an existing cluster $C_k$ and $m_j$ is matched to an existing cluster $C_s$ (or \emph{vice versa}, \ie $m_i$ to $C_s$ and $m_j$ to $C_k$). This implies an assumption, meaning that the assignment of methods to cluster is correct or that, in other words, when a method is assigned to a cluster, the method actually implements functionalities related to those of the cluster. To partially address this threat, two of the authors manually analyzed a set of 100 methods assigned by \approach to a specific cluster, with the goal of verifying whether the assigned cluster actually implements the same feature of the method. 

After solving conflicts arisen in 7\% of cases, they reported an accuracy of 91\%. Thus, we acknowledge possible imprecisions.

\emph{Internal validity.} We tuned the \approach's parameters on a set of commits not used for the learning of the association rules nor for assessment of \approach's performance. We experimented with 1,080 combinations of parameters. However, it is possible that better performance can be achieved by considering other possible values. Thus, from this point of view, the reported performance is an underestimation. We adopted a careful experimental design to avoid using ``data from the future'' when tuning and testing our approach.

\emph{External validity.} Overall, our study involves \totalApps open-source Android apps. The main issue is related to the fact that all used apps are open source, and might not be representative of commercial apps. Also, while \approach is general enough to be adapted to other contexts (\eg Java programming in general), we decided to focus on a more narrow scenario at least for this first work.

%% file: related.tex

\section{Related Work} \label{sec:related}

\approach is one of the many recommender systems proposed in the software engineering literature. The latter have been proposed to support many different tasks, such as the recommendation of formal and informal documentation (see \eg \cite{Treu2016a,Wong2013a,Ponz2017a}), the automatic generation of code for different purposes (\eg \cite{Tufano:ase2018,Tufano:icse2019,Lezos2016,Glass1996,Liao2010,Singh2013}), or the recommendation of relevant code examples/discussions for a task at hand (\eg \cite{Cord2012a,Rigb2013a,Taku2011a,Holm2005a,Holm2006a}). We focus our discussion on the most related works, and in particular on those dealing with code completion techniques and code search engines.

\subsection{Code Completion Techniques}

Basic code completion features of IDEs often rely on the static type system of a programming language and do not consider the actual code context. Suggestions are usually sorted, \eg in alphabetical order. As a result, relevant recommendations are not always easy to identify.

An alternative approach was presented by Bruch \etal \cite{Bruch:fse2009}. Their \textit{intelligent code completion system}
filters out candidates from the list of tokens recommended by the IDE that are not relevant to the current working context, and
ranks candidates based on how relevant to the context they are. 

Another context-sensitive approach was developed by Nguyen \etal \cite{Nguyen:icse2012}. Their \textit{GraPacc} method uses graphs to model API usage patterns, where nodes represent actions (\eg method calls) and control points (\eg while), and edges represent control and data flow dependencies between nodes. Context information such as the relation between API elements and other code elements is considered for ranking the most fitted API usage patterns.

Statistical language models have also been used for code completion. In their seminal work on the naturalness of software, Hindle \etal developed a code completion engine for Java, based on an n-gram language model \cite{Hindle:icse2012}. Their work has been extended by Nguyen \etal \cite{Nguyen:msr2016} and Tu \etal \cite{Tu:fse2014}. 

A language model approach was implemented by Raychev \etal too \cite{Raychev:pldi14}. They extract sequences of method calls from a large codebase to train a model, which they use to support the autocompletion of method calls, achieving an accuracy of 90\% when considering the top three results. Method call completion was also explored by Asaduzzaman \etal \cite{Asaduzzaman2014}. Their approach, called CSCC, relies on a database of method call usage contexts collected from open source projects and applies a hash function to find relevant recommendations. From another perspective, Robbes and Lanza proposed to improve code completion by focusing on the recent changes implemented by the developer \cite{Robb2010a}.

Popular IDEs have recognized the importance of supporting context-sensitive recommendations. For example, IntelliJ IDEA has a feature called \textit{Smart completion} to filter and show suggestions applicable to the current context. NetBeans has a \textit{Smart Code Completion} feature to display at the top of the suggestions the most relevant ones for the context. Eclipse has plugins to extend its core code completion, among these, \textit{aiX Code Completer}\cite{aix} and \textit{Codota}\cite{codota} use AI techniques and can even recommend a full line of code.

While these approaches are undoubtedly valuable to speed up code writing, they are limited to recommendations related to the next few tokens the developer is likely to type given the current context. In the best case, they can recommend a few APIs that the developer is likely to use next. With \approach we forge another step ahead, to predict the next full method a developer is likely to implement.

\subsection{Code Search Engines}

\approach is also related to approaches implementing code search engines that allow retrieving code samples and reusable open source code from the Web.

Early online code search engines (\eg codesearch.google.com, koders.com, and krugle.org) offered keyword-based search and file-level retrieval. These approaches could be improved by considering structural and semantic information of code.
Bajracharya \etal \cite{Bajracharya2006} developed Sourcerer, a code search engine that extracts structural information from the code and stores it in a relational model so it can be queried for code search. It supports queries for control structures, Java types, and micro patterns (\eg implementation of Semaphore). 

Reiss developed an approach to combine code search with transformations to map the retrieved code, to meet user specifications \cite{Reiss2009}. For the searching, it allows the user to specify multiple semantic rules, which also form the basis for the transformations.

Thummalapenta \etal developed an approach to support code search engines with static analysis to return fewer, but more relevant code samples for search queries \cite{Thummalapenta2007,Thummalapenta2007b}. Their primary goal was to support a user in reusing a given API. Later they extend their approach with SpotWeb \cite{Thummalapenta2008} to assist users by detecting hotspots that can serve as starting points for reusing APIs.

API usage was also proposed by McMillan \etal \cite{Grechanik2010,McMillan2012} to return highly relevant matches for a source code search engine. Their approach combines three sources of information to locate relevant software: the textual descriptions of applications, the API calls used inside each application, and the dataflow among those API calls.

Compared to code search engines, \approach also relies on an extensive database of methods' source code in open source applications. These methods are organized in clusters based on a similarity algorithm implemented in the ASIA clone detector \cite{aghajani2019automated}. \approach does not require the user to write a ``query'' to identify relevant pieces of code, but extrapolates this need by monitoring the IDE. 

%% file: conclusion.tex

\section{Conclusions} \label{sec:conclusions}

Code completion, while provenly useful and extensively used by developers \cite{MurphyKF06} is just a step in the direction of an automated pair programmer, adding complete methods that a developer would have to add anyway and thus removing from the developer the burden of rote work. This was the ambitious goal that we set out to achieve with this work, embodied in the creation of \approach, an approach and a tool \cite{replication} to automatically recommend to developers the complete next method to write during implementation activities. 

\approach relies on a simple but intuitive idea: programming is an eclectic activity, which some even go as far as calling it ``natural'' \cite{Hindle:icse2012}. What a developer is doing has a high chance of having been done by someone else, somewhere else before. Leveraging this idea, \approach mines vast amounts of data to recommend complete methods given a set of methods being implemented by a developer. We evaluated \approach on the change history of \totalApps Android apps. The results show the potential of \approach, with hundreds of correct methods recommended even in its most conservative configuration. 

However, our findings are not conclusive for what concerns the actual usefulness of the generated recommendations in a real usage scenario, in which developers use \approach during coding activities. This is due to two observations we made. First, some of the methods recommended by \approach are quite short and, while they can still be useful, they could also represent ``trivial'' recommendation for developers. We believe this can in part be made up by introducing a user feedback loop, which is part of our future work. The quantitative results show that around 15\% of the tokens from the recommendations need to be modified, added or deleted to fit the user's code base. One of our future plans is to integrate code adaption techniques into \approach to avoid potential conflicts or compilation errors with the user's code environment, and convert the coding convention into the user's style. Second, due to our experimental design, the ``unmatched recommendations'' are always considered false positives, while we observed that some are actually valuable recommendations. Thus, a deeper evaluation of \approach including a well-designed user study represents another main target of our future research.